\documentclass[journal=jacsat,manuscript=article]{achemso}
\setkeys{acs}{keywords = true}

\usepackage{chemformula} 
\usepackage[T1]{fontenc} 
\usepackage{amsmath,amssymb,bm}
\usepackage{graphicx}
\usepackage{multirow}

\author{Sabitoj Singh Virk}
\author{Patrick T. Underhill}
\email{underp3@rpi.edu}
\affiliation{Rensselaer Polytechnic Institute, 110 8th St, Troy, New York 12180, USA}

\title{Application of a simple short-range attraction long-range repulsion colloidal model towards predicting the viscosity of protein solutions}


\begin{document}



\begin{abstract}
Some hard sphere colloidal models have been criticized for inaccurately predicting the solution viscosity of complex biological molecules like proteins. Competing short-range attractions and long-range repulsions, also known as SALR interactions, have been thought to affect the microstructure of a protein solution at low to moderate ionic strength. However, such interactions have been implicated primarily in causing phase transition, protein gelation, or reversible cluster formation and their effect on protein solution viscosity change is not fully understood. In this work we show the application of a hard sphere colloidal model with SALR interactions towards predicting the viscosity of dilute to semi-dilute protein solutions. The comparison is performed for a globular shaped albumin and Y-shaped therapeutic monoclonal antibody that are not explained by previous colloidal models. The model predictions show that it is the coupling between attractions and repulsions that give rise to the observed experimental trends in solution viscosity as a function of pH, concentration, and ionic strength. The parameters of the model are obtained from measurements of the second virial coefficient and net surface charge/zeta-potential, without additional fitting of the viscosity.
\end{abstract}

\section{Introduction}\label{sec_intro}
Protein-protein interactions (PPIs) play an integral part in driving many biophysical and biological processes of proteins in solutions, a few of which are phase separation, aggregation, intracellular signalling pathways, and molecular transport. Neurodegenerative diseases like Alzheimer, Parkinson and Creutzfeldt-Jakob are known now to be caused by aberrant PPIs resulting in protein aggregation~\citep{ross_2004, spires_2017, hernandez_2020}. Similarly, although protein crystallization mechanisms are still unclear~\citep{sleutel_2018} but nucleation pathways of crystals have been found to be sensitive to a balance of specific and non-specific protein interactions~\citep{whitelam_2010}.

Protein solutions such as recombinant globular proteins and therapeutic monoclonal antibodies are routinely formulated and subcutaneously administered at high concentrations ($>$ 50 mg/ml) due to potency concerns~\citep{kollar_2020}. Undesirable high viscosity is a common occurrence in high concentration formulations/injections which may even alter or render the drug ineffective~\citep{shire_2004}. Numerous other variables than concentration, related either to the protein bio-molecule (surface charge, shape, molecular and charge anisotropy) or solution conditions (salt concentration, pH, solvent viscosity, temperature) can further affect the protein solution viscosity~\citep{yadav_2010}. Such a broad array of factors that can individually affect the protein-protein interactions have made the problem of understanding or predicting viscosity behavior difficult.

Colloidal hard sphere models have been able to explain many behaviors of protein solutions like aggregation, self-assembly, and stability~\citep{piazza_1998, heinen_2012}. These studies consider the dispersed particles under the influence of a strong single type of particle-particle interaction. Hard sphere models under the assumption of a single and strong dominating interaction have however noticeably failed to explain or predict the viscosity behavior of some protein solutions under variable conditions of pH, protein concentration, and added salt concentration~\citep{Sarangapani_2013, Sarangapani_2015,Binabaji_2015,pathak_2021}. This failure of classical colloidal models in predicting the viscosity of protein solution has been used as a source of criticism on the usefulness of colloidal models towards understanding the dynamic and structural properties of the protein solutions~\citep{pathak_2021,prausnitz_2015,roche_2022}.

Previous work has shown that instead of a single dominant interaction (attraction or repulsion), the combined effect of equally strong competing interactions (short-range attraction and long-range repulsion), can change qualitative trends in a colloidal model of solution viscosity~\citep{tang_2022}. SALR(short-range attraction and long-range repulsion) interactions have experimentally demonstrated their effect in protein solutions at low to moderate salt concentration. Lysozyme, a globular protein solution displays an intermediate range order structure  as well as different phases like dispersed fluid, clustered fluid, random percolated, and glassy states~\citep{stradner_2004,liu_2011,godfrin_2015,godfrin_2018}. Reversible cluster formation has also been observed in many monoclonal antibody solutions~\citep{liu_2005,yearley_2013}. In these cases, the change in microstructure of the protein solution due to the competing interactions has been thought to affect the viscosity significantly. Still, experimental and theoretical studies of SALR systems have focused on understanding microstructural ordering and phase transition of protein solutions~\citep{liu_2019}.

In this paper, we apply our isotropic colloidal model of SALR interactions~\citep{tang_2022} to predict the viscosity behavior of two different classes of protein solutions: a. Globular shaped bovine serum albumin (BSA) protein solution at low ionic strength over a wide range of pH conditions and concentration. b. Y-shaped therapeutic monoclonal antibody (mAb) solution as a function of solution salt concentration, pH, and protein concentration. These molecules are unrelated but share similar trends in their viscosity behavior with solution conditions (indicative of SALR interactions), that currently remains unexplained from the classical colloidal models perspective~\citep{Sarangapani_2013,Binabaji_2015}. Through these examples, our aim is to show that SALR PPIs could be the driving interactions behind the complex viscosity behavior of dilute to semi-dilute protein solutions. Further, coupling between the interactions is important for designing therapeutic protein formulations where low viscosity is desired at high concentrations.
\section{Methods}{\label{sec_methods}}

\subsection{\label{viscosity_model}SALR colloidal model for protein-protein interactions}

We briefly describe the colloidal model for SALR protein interactions developed and analysed in our recent work~\citep{tang_2022}. The long-range electrostatic repulsions were modelled using an approximate screened-Coulomb (SC) potential given by~\citep{russel_1976}
\begin{equation} \label{eqn:Vsc}
    V_{SC}\left(\rho\right)=\alpha\frac{e^{-\kappa\rho}}{\kappa\rho}
\end{equation}
where $\rho$ is the center-to-center distance between two particles nondimensionalized by the particle radius $a$ and $\kappa$ is the inverse of the  Debye length nondimensionalized by the particle radius given by
\begin{equation} \label{eqn:kappa}
   \kappa = a \left( \frac{2N_{A}e^{2}I}{\epsilon_{f}kT} \right)^{1/2}
\end{equation}
where $N_{A}$ is Avogadro's number, $e$ is the elementary charge, $I$ is the ionic strength of the solution, $\epsilon_f$ is the permittivity of the fluid, $k$ is Boltzmann's constant, and $T$ is the absolute temperature.

The dimensionless group $\alpha$ represents the repulsion strength of the particle interactions and is given by
\begin{equation} \label{eqn:alpha}
    \alpha = \frac{4\pi\epsilon_{f}\psi^{2}_{0}a}{kT}\kappa e^{2\kappa}
\end{equation}
where $\psi_{0}$ is the zeta potential. $\alpha$ can also be written in terms of molecular net surface charge $Z$ using Debye-Huckel approximation in conditions where the magnitude of the zeta potential $\psi_{0}$ is less than $\frac{kT}{e}$~\citep{hunter_2013}. The net surface charge $Z$ for a uniformly charged sphere is related to the measured zeta potential as
\begin{equation} \label{eqn:Zapprox}
    Z = \frac{4\pi\epsilon_{f} a \left(1 + \kappa\right)\psi_{0}}{e}
\end{equation}
Substituting equation $\left(\ref{eqn:Zapprox}\right)$ into equation $\left(\ref{eqn:alpha}\right)$ and representing the repulsion strength in terms of net surface charge
\begin{equation} \label{eqn:alpha*}
    \alpha = \frac{(Ze)^{2}}{4\pi\epsilon_{f}kTa(1+\kappa)^{2}}\kappa e^{2\kappa}
\end{equation}
Substitution of equation $\left(\ref{eqn:alpha*}\right)$ into equation $\left(\ref{eqn:Vsc}\right)$ gives the screened Coulomb potential developed by Vilker~\cite{vilker_1981} for small surface potentials ($<$ $\frac{kT}{e}$) that is used here.

The short-ranged attractive interactions were modeled using a Morse potential~\citep{holmes-cerfon_2017} which for $\rho>2$ is
\begin{equation} \label{eqn:morse}
    V_M\left(\rho\right) = \varepsilon_d e^{-\left(\rho-2\right)/\epsilon}\left(e^{-\left(\rho-2\right)/\epsilon}-2\right)
\end{equation}
where $\varepsilon_d$ denotes the attractive well depth at $\rho=2$ and $\epsilon$ is the sticky interaction range. An effective stickiness parameter $\tau^*$ denoting the strength of short-range attractions for this model was calculated using the integral of the equilibrium radial distribution function $g_{M}\left(\rho\right)=e^{-V_{M}\left(\rho\right)}$
\begin{equation} \label{eqn:tauEff}
    \frac{1}{4}\int_{2}^{\infty}\left[g_{M}\left(\rho\right)-1\right]\rho^2d\rho = \frac{1}{6\tau^*}.
\end{equation}
This is equivalent to using the second virial coefficient for the case of sticky hard spheres to define the strength of attractions.

Ref.~\citep{tang_2022} developed analytical approximations of the second virial coefficient and zero-shear viscosity coefficient which will be used in this work. The second-virial coefficient $B_2$ of a dilute suspension can be used to measure the equilibrium protein-protein pair interactions. Theoretically, the $B_2$ is scaled by the hard sphere result $B_{2,HS}=\frac{16\pi a^3}{3}$ to produce $B_2^* = B_2/B_{2,HS}$. For this model $B_2^{*}$ can be approximated as~\citep{tang_2022}
\begin{equation} \label{eqn:B2*analytical}
    B_2^* \approx  1 - \frac{1}{4\tau^*}e^{-V_C}+b_{1}\left(\frac{(e-1)L_0^3}{3e}  \right)
\end{equation}
where $V_C = V_{SC}|_{\rho=2}=\alpha e^{-2 \kappa} /(2 \kappa)$ is the long-ranged electrostatic potential at/near contact and $L_0$ is the dimensionless inter-particle distance at which the electrostatic potential and thermal energy terms balance each other~\citep{russel_1976}. The constant $b_{1}$ in equation $\left(\ref{eqn:B2*analytical}\right)$ was previously inserted to match the approximation with numerical calculations. It was found to be order one, and depend on the dimensionless Debye length. Here we will use it as a fit parameter when determining model parameters from the experimental second virial coefficient. For $\alpha\gg1$, $L_0$ can be approximated as~\citep{russel_1976}
\begin{equation} \label{eqn:L_0}
    L_0 \approx \frac{1}{\kappa}\ln\frac{\alpha}{\ln\left(\alpha/\ln\alpha\right)}
\end{equation}

For a dilute suspension of particles the relative zero-shear viscosity $\eta_{r}$ can be written as a series in the concentration $c$ of the particles (protein)~\citep{russel_1984}
\begin{equation} \label{eqn:etarelative}
    \eta_{r} = \frac{\eta}{\eta_{0}}= 1 + [\eta] c + k_H\left([\eta]c\right)^2 + \cdots
\end{equation}
where $\eta$ and $\eta_{0}$ are the viscosity of the solution and the solvent respectively, $[\eta]$ is the intrinsic viscosity, and $k_H$ is the Huggins coefficient. For a suspension of spheres the intrinsic viscosity is related to the hydrodynamic radius by
\begin{equation} \label{eqn:intrinsicviscosity}
    [\eta] = \frac{10\pi N_{A}a^{3}}{3M_{w}}
\end{equation}
where $M_{w}$ is the molar mass of the object. In situations where the $c^2$ term is important, the fluid can be non-Newtonian, in which the viscosity depends on shear rate and the type of flow. For the experiments used here, a shear flow was used in the measurements and the viscosity was nearly independent of shear rate.

For spheres the relative viscosity of the solution can also be expressed in terms of volume fraction of the solute $\phi$, in which case the quadratic coefficient of viscosity is denoted by $c_2$
\begin{equation} \label{eqn:relativeviscosity}
     \eta_{r} = 1 + 2.5\phi + c_2\phi^2 + \cdots
\end{equation}
By matching the two expressions for $\eta_r$, the Huggins coefficient can be determined from $c_2$ or vice-versa using $k_H=c_{2}/2.5^2$.

This work will use an analytical approximation for $c_2$ given by Ref.~\citep{tang_2022} in terms of the hydrodynamic contribution $c_2^H$, Brownian contribution $c_2^B$, and interaction contribution $c_2^I$ given by
\begin{equation}
\label{eqn:c2analytical}
    \begin{split}
        & c_{2} = c_{2}^{H} + c_{2}^{B} + c_{2}^{I} \\
        & c_{2}^{H} = \frac{5}{2}+e^{-V_C}\left(\frac{1.11}{\tau^{*}}+ 2.03\right)+0.69 \\
         & c_{2}^{B} = 0.96\left[\left(\frac{1}{\tau^*}\right)e^{-V_C}\left(1.37 + \frac{0.21\left(\frac{1}{\tau^*}\right) \ln{\epsilon }}{-0.05\left(\frac{1}{\tau^*}\right)\ln{\epsilon}+1} + \frac{2}{3}\alpha e^{-2\kappa}\right) + 1.01\right] \\
        & c_{2}^{I} = -1.22\left(\frac{1}{\tau^*}\right)e^{-V_C}\left(1.37 + \frac{0.21\left(\frac{1}{\tau^*}\right) \ln{\epsilon }}{-0.05\left(\frac{1}{\tau^*}\right)\ln{\epsilon}+1} + \frac{2}{3}\alpha e^{-2\kappa}\right)+ \\ & A_{1}\left(\frac{3}{40}\right)\left(\frac{1}{\kappa}\ln\frac{\alpha}{\ln\left(\alpha/\ln\alpha\right)}\right)^2{\ln\left(\frac{\alpha}{\ln\alpha}\right)}\left(\frac{1}{6}\left(\frac{1}{\kappa}\ln\frac{\alpha}{\ln\left(\alpha/\ln\alpha\right)}\right)^2{\ln\left(\frac{\alpha}{\ln\alpha}\right)}+1.37\right)
    \end{split}
\end{equation}
The parameter $A_1$ was previously inserted to match the approximation with numerical calculations. It was found to be order one, and depend on the dimensionless Debye length. Here we will use $A_1=1$ as an estimate for the ionic conditions to be analyzed and so that the viscosity calculations involve no fitting of experimental viscosity measurements.

\section{\label{sec_results_discussion}Results and Discussion}
The results here focus on viscosity measurements in the literature that were not previously explained by simple colloidal models of viscosity. The work of Sarangapani~\citep{Sarangapani_2013} measured  bovine serum albumin (BSA) protein solution viscosity at low ionic strength over a wide range of concentration (2 mg/ml $\leq$  c $\leq$ 400 mg/ml) and concurrently varying the pH (4.0 $\leq$ pH $\leq$ 7.4). We simultaneously apply our approximations to predict the viscosity behaviour of a therapeutic monoclonal antibody (mAb) solution in a diverse buffer pH (5.0 $\leq$ pH $\leq$ 7.0) at a concentration range (0 mg/ml $\leq$  c $\leq$ 250 mg/ml) and with NaCl concentrations ranging from 13mM to 103mM~\citep{Binabaji_2015}. The experimental data presents a broad parameter space so are an ideal system to understand the application of our SALR colloidal model. We do so by determining first the parameters using the charge and the second-virial coefficient then compare the model predictions with the experimental viscosity without additional fitting. The highest concentrations correspond to volume fractions $\phi \sim 0.6$. It is expected that the quadratic expansion used will not be quantitatively accurate at these concentrations, but will be accurate at intermediate concentrations.

\subsection{\label{protein_second-virial coefficient}Determining PPI parameters from second-virial coefficient}

The experimental net surface charge on the BSA molecules at different pH was taken from the hydrogen ion titration data of Ref.~\citet{tanford_1956}. The surface charges  are in agreement with other BSA solution studies~\citep{zhang_2007,li_2016}. The effective hydrodynamic/Stokes radius $a$ of a BSA monomer molecule at all the pH samples was taken to be 3.4 nm which is an experimentally determined value of stokes radius~\cite{gonzalez_2003} and agrees well with other experimentally reported values~\citep{kuntz_1974,axelsson_1978}. The ionic strength of 20 mM at 25$^{\circ}$C used gives a $\kappa= 1.59$. Together these produce the $\alpha$ values given in Table \ref{table:1}.

\begin{table}
\centering
\begin{tabular}{ |p{3cm}|p{3cm}|p{3cm}| }
 \hline
 \multicolumn{3}{|c|}%
{Bovine Serum Albumin (BSA)}\\%
 \hline
 \multicolumn{3}{|c|}%
{pI=4.8; $M_{w}$=67 kDa; a=3.4 nm}\\%
 \hline
 pH & Z & $\alpha$ \\
 \hline
 4.0   & +8  &76.8\\
 \hline
 5.0   & -4  &19.2 \\
 \hline
 6.0   & -9  &97.3 \\
 \hline
 7.4   & -12 &172.9 \\
 \hline
\end{tabular}

\caption{PPI parameters obtained for BSA solution using the net surface charge Z at different pH from Ref.~\citet{tanford_1956}.}
\label{table:1}
\end{table}

For determining the short-range attraction strength ($1/\tau^*$) of BSA solution, we fit the analytical approximation of $B_2^*$ using equation $\left(\ref{eqn:B2*analytical}\right)$ to the experimentally determined values from Ref.~\cite{Sarangapani_2013}. This fitting, along with using $M_{w}$=67 kDa, results in a value of $\tau^*=0.043$ along with $b_1=0.8$. Figure 1 shows a comparison of the second virial coefficient between the colloidal model and experiments. At pH 5.0, the small surface charge maximizes the contribution from short-ranged attractions. The fit using the model matches both this negative $B_2$ as well as the positive $B_2$ for other values of pH.

\begin{figure}
\centering
\includegraphics[scale=0.9]{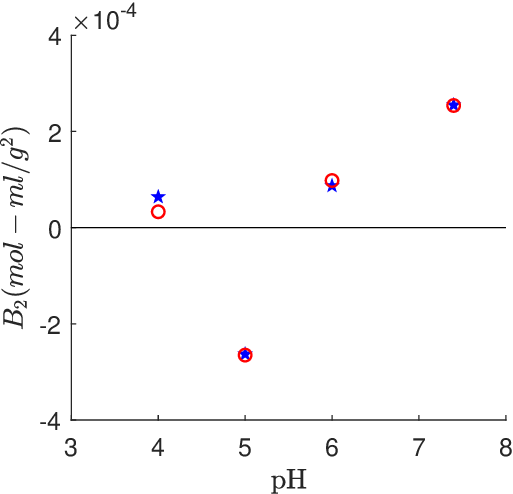}

    \caption{ BSA second virial coefficient $B_2$ as a function of solution pH. The blue pentagrams represents the static light scattering (SLS) experiments performed at an ionic strength of 20 mM and 298.15 K~\citep{Sarangapani_2013}. The circles
    are the analytical approximation from equation $\left(\ref{eqn:B2*analytical}\right)$ with $\tau^*=0.043$ and $b_{1}=0.8$.}
    \label{fig:01}
\end{figure}

A similar approach is used for an antibody solution based on the system in Ref.~\cite{binabaji_2013} with $M_{w}$=142 kDa and $a=5.43$ nm. The charge was measured for different conditions of pH and ionic strength. The parameters in the colloidal model have been computed for these conditions, and are given in Table 2.

\begin{table}
\centering
\begin{tabular}{ |p{2cm}|p{3cm}|p{2cm}|p{2cm}|p{2cm}| }
 \hline
 \multicolumn{5}{|c|}%
{Monoclonal Antibody (mAb)}\\%
 \hline
 \multicolumn{5}{|c|}%
 {pI=8.1; $M_{w}$=142 kDa; a=5.43 nm}\\%
 \hline
 pH & Ionic strength &Z &$\kappa$ & $\alpha$ \\
 \hline
 5.0 & 13 mM  & 12 &2.04 &248.1 \\
 \hline
 5.0 & 23 mM  & 14 &2.72 &1171.6 \\
 \hline
 5.0 & 103 mM  & 24 &5.75 & 947000 \\
 \hline
 6.0 & 20 mM  & 6 &2.53 &152 \\
 \hline
 7.0 & 20 mM  & 3 &2.53 &38 \\
 \hline
\end{tabular}

\caption{PPI parameters obtained for mAb solution using the net surface charge Z at different pH from Ref.~\citet{binabaji_2013}.}.
\label{table:2}
\end{table}

The short-range attraction strength ($1/\tau^*$) was computed by fitting the $B_2$ from self-interaction chromatography (SIC) measurements. The fitting results in a value $\tau^*=0.031$ and a value $b_1=3.5$. Figure 2 compares the $B_2$ from SIC measurements at $pH=5.0$ with the fit from the colloidal model. The model matches the experiments to within the experimental error bars.

\begin{figure}
    \centering
\includegraphics[scale=0.9]{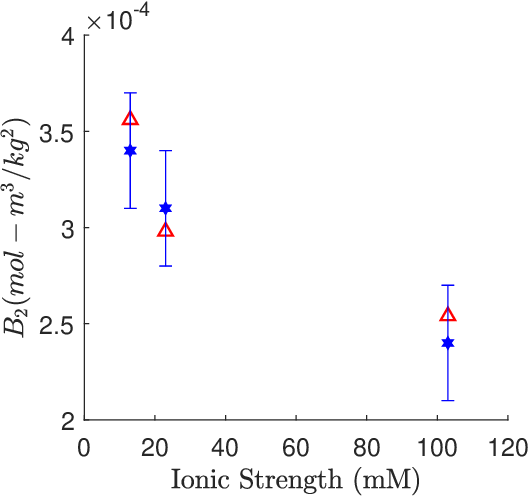}

    \caption{ Therapeutic monoclonal antibody (mAb) second virial coefficient $B_2$ as a function of solution ionic strength. The blue hexagrams represents the measurements with error bars from SIC experiments  at pH 5.0 and 298.15 K~\citep{binabaji_2013}. The triangles are the analytical approximation from equation $\left(\ref{eqn:B2*analytical}\right)$ with $\tau^*=0.031$ and $b_{1}=3.5$.}
    \label{fig:02}
\end{figure}

\subsection{\label{protein_viscosity}Predicting viscosity of protein solutions from PPI parameters}

\subsubsection{\label{pH effect}Effect of pH on the viscosity of protein solutions}

The solution pH can alter the viscosity of a protein solution by changing the net charge on the proteins. A simple model that attempts to capture this effect is a colloidal model of $c_2$ for hard spheres with strong long-range electrostatic repulsions only ($\alpha\gg 1$)~\citep{russel_1976}. For these conditions, the strong repulsions create an excluded region around each sphere, in which the radius of the excluded shell depends on the charge. The quadratic coefficient of viscosity for this screened Coulomb model is given by
\begin{equation} \label{eqn:c2ES}
    c_{2,SC}=\frac{5}{2}+\left(\frac{3}{40}\right)\left(\frac{1}{\kappa}\ln\frac{\alpha}{\ln\left(\alpha/\ln\alpha\right)}\right)^5
\end{equation}
For semi-dilute to concentrated protein solutions, this model fails to predict the viscosity and its behaviour with a pH change~\citep{Sarangapani_2013,Binabaji_2015,saluja_2006,yadav_2011}. The reported viscosity in these studies are the highest at a pH close to the pI (isoelectric point) and lower at pH values where the absolute net charge on the molecule is higher compared to pI. This viscosity behaviour is opposite of what equation (\ref{eqn:c2ES}) would predict, in which a larger $\alpha$ gives a larger $c_2$ and larger viscosity.

Here we will compare the experimental observations with predictions from the colloidal model reviewed earlier that incorporates short ranged attractions. Ref.~\citep{tang_2022} showed that attractions can play an important role in the viscosity behavior even if the $B_2$ is positive (i.e. with respect to osmotic pressure, repulsions are more important than attractions). BSA solution at 300 mg/ml~\citep{Sarangapani_2013} and mAb solution at 150 mg/ml~\citep{Binabaji_2015} were chosen as the example cases for the comparison.

The parameters for computing $c_2$ in equation (\ref{eqn:c2analytical}) and subsequently $\eta_r$ in equation (\ref{eqn:etarelative}) using $k_H=c_{2}/2.5^2$ were taken from Section 3.1. For the given hydrodynamic radii, the solution intrinsic viscosity [$\eta$] (calculated using equation $\left(\ref{eqn:intrinsicviscosity}\right)$) are $[\eta]_{BSA}$ = 0.0037 ml/mg and $[\eta]_{mAb}$ = 0.0071 ml/mg. The parameter $\epsilon$ is the range of the short ranged attractions nondimensionalized by the particle radius. Hydrophobic interactions typically occur over a length scale of a few water molecules. Comparing this to the radii of the proteins examined here, we choose $\epsilon=0.1$ to be in the correct order of magnitude. The weak logarithmic dependence on $\epsilon$ in equation (\ref{eqn:c2analytical}) means that the order of magnitude is the primary concern and it is not necessary to use a different $\epsilon$ for the different proteins.

The expressions in equations (\ref{eqn:morse}) and (\ref{eqn:tauEff}) provide a relationship among $\epsilon$, $\varepsilon_d$, and
$\tau^*$. Therefore, the choice of $\epsilon=0.1$ leads to a one-to-one relationship between $\varepsilon_d$ and $\tau^*$. The value of $\tau^*=0.043$ for BSA corresponds to $\varepsilon_d=3.7$. The value of $\tau^*=0.031$ for the mAb corresponds to $\varepsilon_d=4.0$.

\begin{figure}
   \centering
\includegraphics[scale=0.75]{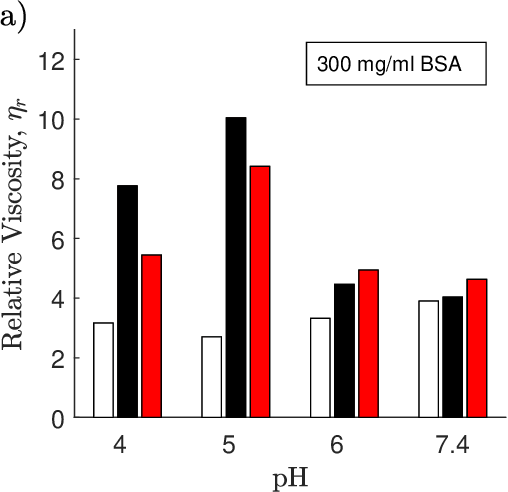}
\includegraphics[scale=0.75]{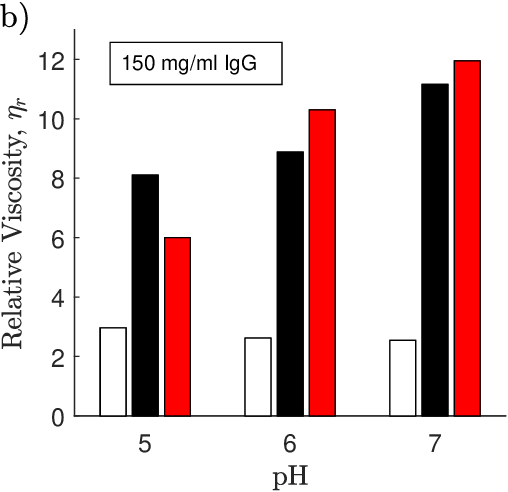}

    \caption{The relative viscosity of a) BSA and b) monoclonal antibody (mAb) solution as a function of solution pH. The black bars show the experimental measurements (BSA~\cite{Sarangapani_2013}; mAb~\cite{Binabaji_2015}). The colloidal model predictions with SALR interactions (equation $\left(\ref{eqn:c2analytical}\right)$) are shown in red bars. The model with only repulsive interactions (equation $\left(\ref{eqn:c2ES}\right)$) are shown in white bars.  }
    \label{fig:03}
\end{figure}

The results of the relative viscosity comparison are shown in Figure \ref{fig:03}. We also show the predictions based on equation (\ref{eqn:c2ES}) that include repulsions only. The model including both attractions and repulsions captures the experimental trends with pH. The model including attractions gives a higher viscosity closer to the isoelectric point, while the model with only repulsions gives a lower viscosity closer to the isoelectric point. The attractions play an important role even though the $B_2$ is positive in most conditions.

The mechanism for this trend stems from the factor $e^{-V_c}$ in equation (\ref{eqn:c2analytical}). This term modulates the increase in viscosity due to attractions via a Boltzmann factor containing the Coulumb potential at contact between two proteins. The repulsions of objects with a net charge reduces the likelihood that two objects get close together where the attractions can increase the viscosity.

\subsubsection{\label{concentration effect}Protein solution viscosity as a function of protein concentration}

\begin{figure}
   \centering
\includegraphics[scale=0.7]{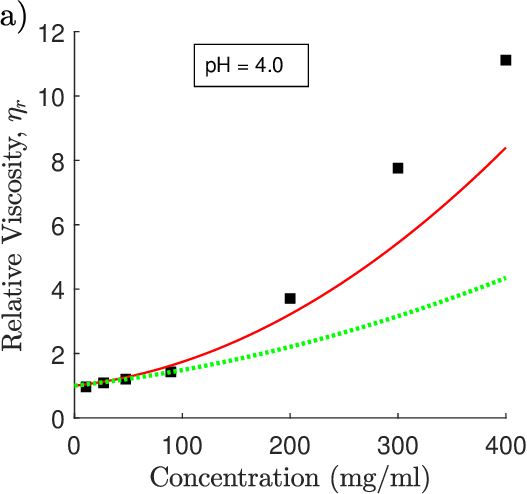}
\includegraphics[scale=0.7]{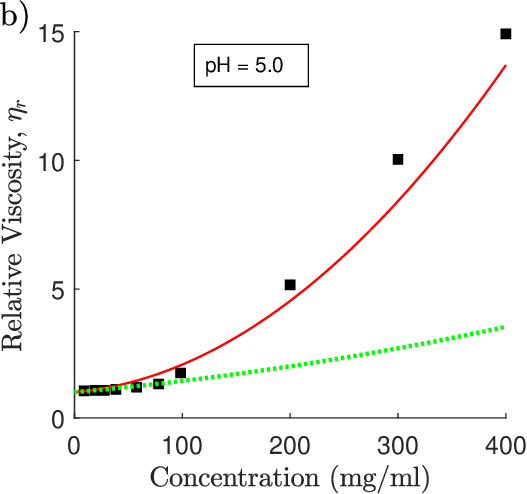}
\includegraphics[scale=0.7]{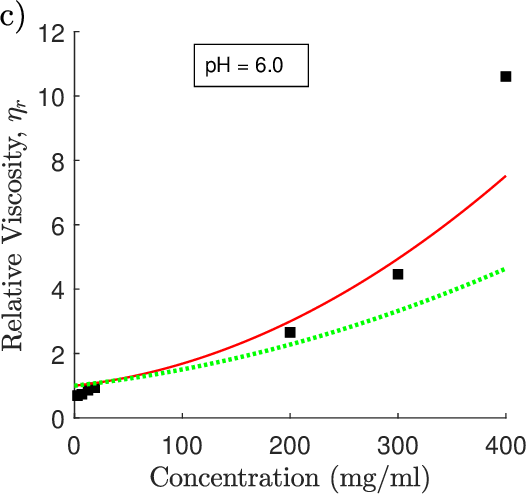}
\includegraphics[scale=0.7]{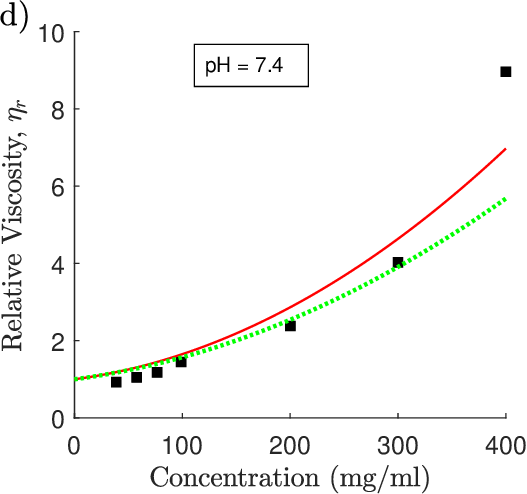}
    \caption{BSA solution relative viscosity $\eta_{r}$ vs. protein concentration at 298.15 K and an ionic strength of 20 mM over a range of solution pH conditions: a) pH = 4.0; b) pH = 5.0; c) pH = 6.0; d) pH = 7.4. The black squares represent the experimental measurements of BSA relative viscosity from the work of Sarangapani~\citep{Sarangapani_2013} while the red solid curve represents the analytical approximation of SALR protein-protein interactions (see equation (\ref{eqn:etarelative}) and (\ref{eqn:c2analytical})). The green dotted line represents the colloidal model of purely repulsive protein-protein interactions (see equation (\ref{eqn:c2ES})).}
    \label{fig:04}
\end{figure}

\begin{figure}
   \centering
\includegraphics[scale=0.7]{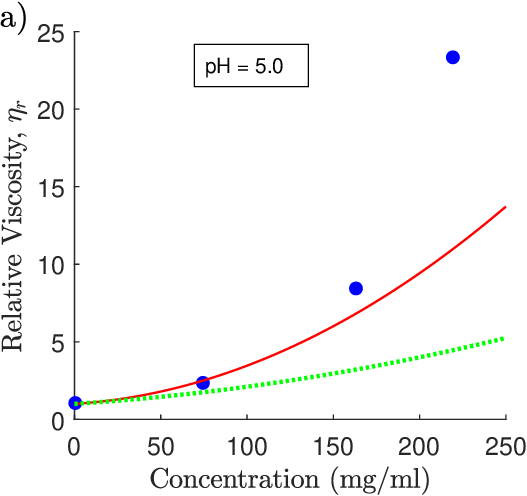}
\includegraphics[scale=0.7]{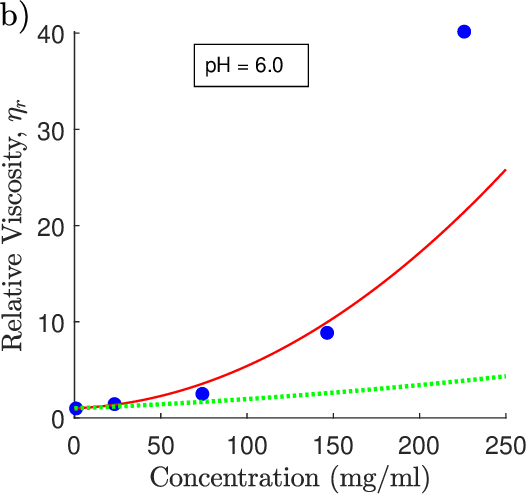}
\includegraphics[scale=0.7]{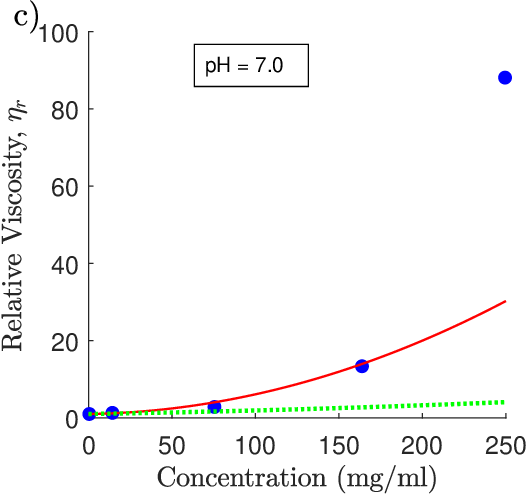}

    \caption{Therapeutic monoclonal antibody (mAb) solution relative viscosity $\eta_{r}$ vs. protein concentration at 298.15 K and a salt concentration of 20 mM over a range of solution pH conditions: a) pH = 5.0; b) pH = 6.0; c) pH = 7.0. The blue circles represents the experimental measurements of antibody solution relative viscosity from the work of Binabaji~\citep{Binabaji_2015} while the red solid curve represents the analytical approximation of SALR protein-protein interactions (see equation (\ref{eqn:etarelative}) and (\ref{eqn:c2analytical})). The green dotted line represents the colloidal model of purely repulsive protein-protein interactions (see equation (\ref{eqn:c2ES})).}
    \label{fig:05}
\end{figure}

The previous comparison as a function of pH focused on a single value of the protein concentration. Since the viscosity of a protein solution can depend strongly on concentration, we show here the comparison between experiments and predictions of the model as a function of protein concentration. Figure \ref{fig:04} shows the comparison for BSA solutions while Figure \ref{fig:05} shows the comparison for mAb solutions. The experimental data were again extracted from the work of Sarangapani~\citep{Sarangapani_2013} and Binabaji~\citep{Binabaji_2015}. The predictions of the colloidal model including attractions are compared with the model only including repulsions.

At very low concentrations, the viscosity is linear in concentration. In this region, the viscosity is independent of protein-protein interactions, so both models are the same and match well with the experimental data. This suggests that the protein radii $a$ that are used (and the resulting intrinsic viscosity) are reasonable.

At higher concentrations the viscosity increases quadratically with a quadratic coefficient that depends on protein-protein interactions. Without additional fitting, the colloidal model including attractions matches the experiments nearly quantitatively. The model that only includes repulsions underpredicts the experiments except for the higher pH BSA solutions for which the repulsions are stronger.

At the highest protein concentrations, the viscosity increases faster than a quadratic, so are not captured quantitatively by a model that truncates the series in equation (\ref{eqn:etarelative}) at the quadratic term. For the data shown here, that typically occurs for concentrations greater than 200 mg/mL, but the details of the deviation from a quadratic depends on the protein and solution conditions. The BSA solution at pH of 5.0 is captured nearly quantitatively up to concentrations of 400 mg/mL.

\subsubsection{\label{IS effect}Effect of ionic strength on the viscosity of protein solutions}

Many studies present the observation that with an increase in the ionic strength, viscosity of an antibody solution is reduced that hints to the dependence of viscosity on electrostatic repulsions which are screened at higher ionic strengths~\citep{liu_2005,kanai_2008,salinas_2010,yearley_2014}. A contrasting and interesting result has been highlighted by other studies where an increase in ionic strength led to an increase in solution viscosity~\citep{Binabaji_2015,meyer_2009,neergaard_2013,lilyestrom_2013}.

\begin{figure}
    \centering
\includegraphics[scale=0.9]{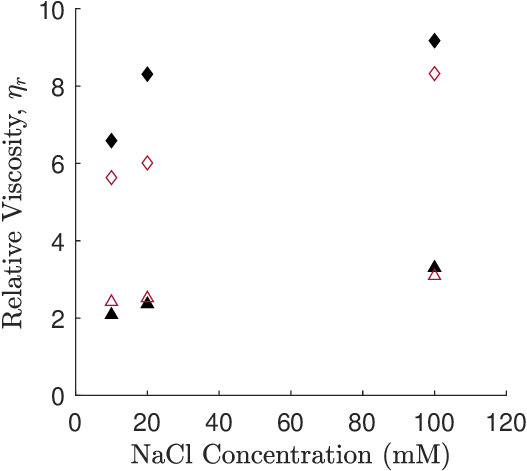}

    \caption{  The relative viscosity of the monoclonal antibody (mAb) solution ($\triangle$, 75 mg/ml; $\Diamond$, 150 mg/ml) as a function of solution NaCl
concentration. The filled symbols are the experiments conducted at pH 5.0 and 298.15 K~\citep{Binabaji_2015}. The empty symbols are the analytical approximation using equation $\left(\ref{eqn:c2analytical}\right)$.}
    \label{fig:06}
\end{figure}

Figure \ref{fig:06} compares the experimental relative viscosity from Ref.~\citep{Binabaji_2015} with the predictions of the colloidal model used here. At 75 mg/mL concentration, the model matches with the experiments. At 150 mg/mL concentration, the model based on a quadratic dependence on concentration underpredicts the experiments similar to the behavior seen in Figure \ref{fig:05}. Importantly the models at both concentrations match the experimental trend of increasing viscosity with increasing salt concentration. This supports the conclusion that protein-protein attractions play an important role in determining the viscosity of the solutions even though the second virial coefficient (shown in Figure \ref{fig:02}) is positive.

\section{\label{sec_conclusion}Conclusion}

We show through this work the application of a colloidal hard sphere model with SALR (short-range attraction and long-range repulsion) interactions in predicting the viscosity of dilute to semi-dilute protein solutions. The predictions have been done to explain the previously published viscosity behaviour of BSA solution~\citep{Sarangapani_2013} and a therapeutic mAb IgG1 solution~\citep{Binabaji_2015}. Both of these studies stand out due to the inability of classical colloidal models in explaining their viscosity trends. We use independent measurements of charge and second virial coefficient to determine the model parameters and predict the viscosity as a function of pH, protein concentration, and ionic strength. The model predictions are nearly quantitatively accurate up to protein concentrations of 150-200 mg/mL or even higher in some cases. This is the case despite the relative simplicity of using a model based on spheres with isotropic interactions and using a quadratic expansion of viscosity in terms of concentration. Through this paper we also showed that in protein solutions of low to moderate ionic strength, SALR interactions may determine the solution viscosity when the repulsions are strong enough to give positive second virial coefficient and prevent a phase transition. This may help enable control over the viscosity of protein solutions through mutations designed to alter the protein attractions and repulsions. It may also help in the understanding of the role of excipients in solution. Some excipients alter short-ranged hydrophobic interactions, which could be approximated by changing the parameter $\tau^*$. Excipients may also lead to slip of water over the protein surface, which would alter the hydrodynamic interactions used in the derivation of equation (\ref{eqn:c2analytical}). In some cases of protein shape and directional interactions, this model will likely not be accurate. But it can then form the basis for comparison of extended colloidal models that include nonspherical objects, anisotropic interactions, and multibody interactions.

\begin{acknowledgement}

This work was supported by the National Science Foundation under Grant No. [CBET 1803497].

\end{acknowledgement}


\providecommand{\noopsort}[1]{}\providecommand{\singleletter}[1]{#1}%
\providecommand{\latin}[1]{#1}
\makeatletter
\providecommand{\doi}
  {\begingroup\let\do\@makeother\dospecials
  \catcode`\{=1 \catcode`\}=2 \doi@aux}
\providecommand{\doi@aux}[1]{\endgroup\texttt{#1}}
\makeatother
\providecommand*\mcitethebibliography{\thebibliography}
\csname @ifundefined\endcsname{endmcitethebibliography}
  {\let\endmcitethebibliography\endthebibliography}{}

\end{document}